\documentclass[prl,preprintnumbers,twocolumn]{revtex4}
\usepackage{graphicx,t1enc}
\usepackage{epsfig}

\usepackage{amssymb,amsmath,amsthm}

\begin{document}

\title{Cultural propagation on social networks}
\author{M. N. Kuperman}
\affiliation{Centro At{\'o}mico Bariloche and Instituto Balseiro,
8400 S. C. de Bariloche, Argentina \\
Consejo Nacional de Investigaciones Cient{\'\i}ficas y
T{\'e}cnicas, Argentina}

\begin{abstract}

In this work we present a model for the propagation of culture on
networks of different topology and by considering different
underlying dynamics. We extend a previous model proposed by
Axelrod by letting a majority govern the dynamics of changes. This
in turn allows us to define a Lyapunov functional for the system.

\end{abstract}

\maketitle

\section{Introduction}
During the last years it has been possible to witness the
increasing interest in mathematical models aimed to describe and
analyze social processes. A fruitful symbiosis took place,
establishing collaborations among researchers from both the social
sciences as well as  physics and mathematics. The process gave
birth to an interesting collection of works dealing with a wide
spectra of social phenomena thus analyzed through a variety of
mathematical and physical techniques \cite{weid,watts,axel1}.

Some important and primordial questions constitute the motivation
to work on these models. It is important to know how the behavior
of unorganized individuals within a society contributes to produce
collective social phenomena. Next, it is necessary to know how
stable this emergent phenomena are. Another important aspect is
knowing to what level social organization is encoded within the
topology of the system, or what is the extent of the effect of the
social structure on the particular characteristics of the
evolution of a given social scenario. A preponderant role
regarding this last aspect has been played by the social or
complex networks.

Proposing different social networks as schemes for the underlying
architecture of the society, many authors have presented models to
describe opinion formation \cite{weid,snaj} , rumors \cite{zan},
diseases \cite{kup, romu} and fashion propagation
\cite{jass,kup2}, urban segregation \cite{hezi}, majority vote
\cite{stau}, etc. In this work we will analyze a generalization of
a model of culture propagation introduced by Axelrod
\cite{axel1,axel}. In the original model, the cultural background
of an individual is characterized by a set of $F$ dynamical
attributes or cultural features, that evolve according to the
social environment. Each feature, in turn, can take $q$ different
values, representing possible traits. The individuals are located
on top of a regular network and interact with their neighbors.
Through this interaction the cultural profile of each individual,
and thus the configuration of the system, evolve. The interaction
is mediated by what is called cultural affinity.  The more similar
an individual is to one of its neighbors, the more likely it is
that they will interact. The interaction consists in the adoption
of a common trait in one of the $F$ cultural aspects.  Typically,
the system evolves towards a monocultural state, but for some
parameters values it freezes in a multicultural state with
coexisting spatial domains of different cultures. The number of
these domains is taken as a measure of cultural diversity.

A systematic analysis of the dependence on $q$ of the original
model was carried out in \cite{cast}. Further analysis of the
model and the role of noise was perform in \cite{palm2,palm1}. In
this work we have modified the original model, allowing for a
wider and more thorough evaluation of the cultural environment
that envelopes each individual and thus influences its cultural
tendencies. Each individual will evaluate whether or not to copy
one of the traits adopted by one of its neighbors, basing the
decision on an observation of the state of its entire
neighborhood. A sort of majority rule will govern the dynamics of
the system. We propose different forms for this majority rule. In
each case, a function of the state of the system that behaves
monotonically in time is found. This function can be associated
with a Lyapunov function of the system. The monocultural state is
always  the absolute minimum of this function, though there exist
some local minima, corresponding to multicultural situations where
the system remains frozen.

\section{The model}

Our work will be based on a previous model by Axelrod
\cite{axel1,axel}. We consider that the cultural background of any
individual can be characterized, in a quite reductionist way, by a
given set of non overlapping features. We will call $F$ the number
of these features $\phi_k$ that define the culture in a schematic
way. In principle, each feature can be associated to a different
aspect of the culture such as spoken language, preferred foods, or
musical style, readings, sports, etc. In turn, a further
subclassification of each feature into categories will serve to
denote the different preferences or traits of the individuals
regarding a particular aspect of his/her cultural background. The
simplest consideration to achieve such subdivision is to consider
that any cultural feature may take on any of $q$ different values,
the same for any $\phi_k$. In principle the values are only
labels, so despite the fact that we will use numbers for
classification, any set of symbols would work as well. Thus, an
individual $i$ will be culturally characterized by a {\it cultural
vector} of $F$ components $\phi^i_k$, each one adopting values
ranging between 1 and $q$. The way to culturally compare two
individuals $i$ and $j$ is to measure their cultural overlap,
$\omega_{(i,j)}$ as follows,
$$\omega_{(i,j)}= \sum_{k=1}^F \delta_{\phi^i_k \phi^j_k}. $$

The individuals will be situated on the vertices of a graph. Two
of them will be considered to be neighbors if they are linked by a
edge. In the original model, the underlying network was a
bidimensional regular lattice \cite{axel}. The topology of the
underlying networks was later generalized in \cite{palm2},
considering amongst others, Small World Networks (SW)
\cite{watts}. In this work, we will consider SW networks as well.

Considering that the individuals will interact with the set of
their neighbors, it will be useful to define at this point the
local overlap of a given individual, $\Omega_i$ as
$$\Omega_i= \sum_{j\in\sigma_i} \omega_{(i,j)},$$
where $\sigma_i$ is the set of neighbors of $i$.
We will also define the quadratic local overlap as
$$\Omega^2_i= \sum_{j\in\sigma_i} \omega_{(i,j)}^2.$$

Starting from an initial distribution of cultural vectors, the
individuals  evolve by analyzing and interacting with their
environment, adapting their cultural preferences according to the
tendencies of the neighborhood. The original numerical simulations
proceed as follows in \cite{axel,palm2,palm1}.  At time step $t$,
a randomly chosen individual $i$ and one of its neighbors $j$ are
evaluated. Their cultural overlap is calculated to decide whether
they will interact or not. The interaction takes place with a
probability $\omega_{(i,j)}(t)/F$, in which case one of the
features $\phi^i_k$ such that $\phi^i_k \neq \phi^j_k$ is set
equal to $\phi^j_k$. Though it is evident that
$\omega_{(i,j)}(t)\geq\omega_{(i,j)}(t-1)$,  the interaction may
affect as well the overlaps between $i$ and the rest of the
neighborhood and thus the change on $\Omega_i$ can not be
anticipated.

Interesting results were obtained in \cite{palm1} by considering
the whole process of cultural dissemination as an optimization
problem. In that work, by analyzing a one dimensional system with
interaction amongst the first neighbors, the authors found a
Lyapunov potential that allowed them to analyze the stability of
the states at which the system remained frozen after some
evolution time. We are interested in extending those results to
more general situations, namely SWN and other complex networks.
The global overlap $$\Omega={1\over 2} \sum_i \Omega_i$$ can not
be claimed to have a monotonic behavior when the system, as
defined, evolves in time. Suppose that as a result of the
interaction between $i$ and $j$ at time $t$ there is a change in
the value of $\phi^i_k$. We will call $\sigma_i^m$ the
neighborhood of $i$ such that for any of the $\mu$ individuals
$h_m\in \sigma_i^m$, $\phi^{h_m}_k(t)=\phi^i_k(t)$; and
$\sigma_i^n$ the set of $\nu$ neighbors $h_n$ such that
$\phi^{h_n}_k(t)=\phi^i_k(t+1)$. The rest of the  of neighbors
will be included in the set $\sigma_i^l$. If at each time step
only one change is allowed, we can calculate
$\Delta\Omega=\Omega(t+1)-\Omega(t)=2(\Omega_i(t+1)-\Omega_i(t))$
by considering that
\begin{eqnarray}
\Omega_i(t+1)&=&\sum_{j\in\sigma_i^l}(\omega_{(i,j)}(t))+ \\
&&\sum_{j\in\sigma_i^m} (\omega_{(i,j)}(t)-1)
+\sum_{j\in\sigma_i^n} (\omega_{(i,j)}(t)+1) \nonumber
\end{eqnarray}
$$=\sum_{j\in\sigma_i}(\omega_{(i,j)}(t))-\sum_{j\in\sigma_i^m}
1 +\sum_{j\in\sigma_i^n} 1.$$ Thus
\begin{equation}
\Omega_i(t+1)=\Omega_i(t)+\nu-\mu.
\end{equation}
The change in $\Omega$ is thus $\Delta \Omega=\nu-\mu$, which is
not necessarily equal or greater than zero. By introducing a
modification in the dynamics of the original model we can assure
that this condition will be fulfilled and thus can think of a
Lyapunov functional for the system.

We will consider different types of dynamics, each one associated
with a corresponding Lyapunov function but at the same time with a
clear social interpretation of the behavior of the individuals.

The underlying network will be built up following the procedure
described in \cite{ws}. In the original model of SW networks, a
single parameter $p$, running from 0 to 1, characterizes the
degree of disorder of the network, respectively ranging from a
regular lattice to a completely random graph. The construction of
these networks starts from a regular, one-dimensional, periodic
lattice of $N$ elements linked to $2K$ nearest neighbors. Then
each of the sites is visited, rewiring $K$ of its links with
probability $p$. Values of $p$ within the interval $[0,1]$ produce
a continuous spectrum of small world networks.

\section{Cultural exchange dynamics}

\subsection{Case 1: Restricted cultural affinity}
The original model proposed by Axelrod considered a very special
case of  biased dynamics for the interactions of the individuals.
Despite the fact that individuals are immersed in their
neighborhood, this  was ignored by requiring that the individual
interact with only one of its neighbors. Taking this fact into
account, a first adaptation of the original model consists in
deciding whether to change or not the value of the chosen feature
by weighting the decision with a further evaluation of the
influence of the neighborhood. Given that the individual $i$
interacts with $j$, the possibility of adopting $\phi^j_k$ for
$\phi^i_k$ will depend on the result of an evaluation following a
sort of majority rule. If by accepting the change of the value of
$\phi^i_k$ $i$ will share the value taken by $\phi^i_k$ with a
bigger group than if by rejecting the change,  then  $i$ accepts
the change. With a probability 1/2 the change is accepted in case
of equality. This is translated into the following situation. The
change is accepted whenever $\nu
>\mu$, and with probability 1/2 when $\nu = \mu$ . Under this
condition, $ \Omega(t+1)-\Omega(t)\geq 0$. So, we will take
$\pounds_1(t)=-\Omega(t)$ as the Lyapunov function of this
dynamics, and the system will evolve to reach a local or absolute
minimum.

\subsection{Case 2: Complete Cultural affinity }

The former rule assures that the global cultural background grows
or at least is maintained constant, while the individuals knows
that accepting the change will warrant being in bigger group
regarding the changing feature. But basing the decision of the
individual on the comparison of only one feature seems quite
myopic. We can propose another condition, making the individual
base the decision on a further evaluation of the local partial
overlaps, $\Omega_i^{m}=\sum_{j\in\sigma_i^m} \omega_{(i,j)} $ and
$\Omega_i^{n}=\sum_{j\in\sigma_i^n} \omega_{(i,j)}$. Despite that
one feature was chosen to be changed, the individual decides
whether to adopt the new value or not by weighting the whole
cultural overlap with its neighborhood and not by analyzing what
happens with the specific feature to be  changed. This is
equivalent to saying that the individual will favor a majority
weighted by deeper cultural affinity. Now, we can no longer say
that $ \Omega(t+1)-\Omega(t)\geq 0$. We must look for another
quantity $\pounds_2(t)$, such that
$\pounds_2(t+1)-\pounds_2(t)\leq 0$. In what follows we show that
$\pounds(t)_3=-(\Omega^2(t) + \Omega(t))$, with
$\Omega^2(t)={1\over 2}\sum_i \Omega^2_i(t)$, satisfies the
required condition.

Let us consider that in the proposed interaction between $i$ and
$j$, $\phi^i_k$ will be change by $\phi^j_k$. There are three
classes of individuals among the neighbors of $i$, those belonging
to $\sigma_i^m$, those belonging to $\sigma_i^n$ and the rest,
that will be grouped in $\sigma_i^l$. The local partial overlap
$\Omega_i^{l}=\sum_{j\in\sigma_i^l} \omega_{(i,j)} $ will no be
affected, regardless of whether or not the interaction takes
place. On the contrary, if the interaction occurs at time $t$,
$\Omega_i^m(t+1)=\Omega_i^m(t)-\mu$ and
$\Omega_i^n(t+1)=\Omega_i^n(t)+\nu$.

If no interaction is allowed, $\pounds_2(t+1)=\pounds_2(t)$. If on
the contrary, the change is accepted, we have
$$\pounds_2(t+1)-\pounds_2(t)=[\Omega^2_i(t+1)-\Omega^2_i(t)]+[\Omega_i(t+1)-\Omega_i(t)].$$
We can expand the rhs of the former equation by considering sums
over $\sigma_i^m$, $\sigma_i^n$, and $\sigma_i^l$. On one side we
have
\begin{eqnarray}
\Omega^2_i(t+1)&=&\sum_{j\in\sigma_i^l}(\omega_{(i,j)}(t))^2+\\
&&\sum_{j\in\sigma_i^m} (\omega_{(i,j)}(t)-1)^2
+\sum_{j\in\sigma_i^n} (\omega_{(i,j)}(t)+1)^2 \nonumber
\end{eqnarray}
$$=\sum_{j\in\sigma_i}(\omega_{(i,j)}(t))^2+\sum_{j\in\sigma_i^m} 1-2 \omega_{(i,j)}(t)
+\sum_{j\in\sigma_i^n} 1+2 \omega_{(i,j)}(t) $$ Expanding and
regrouping terms we get
$$\Omega^2_i(t+1)=\Omega^2_i(t)+\Omega_i^n(t+1) - \Omega_i^m(t+1)+
\Omega_i^n(t) - \Omega_i^m(t),$$
$$=\Omega^2_i(t)+2(\Omega_i^n(t+1) - \Omega_i^m(t))+
\mu - \nu$$ On the other hand we have Eq. 2. Finally
$$\pounds_2(t+1)-\pounds_2(t)=-(2(\Omega_i^n(t+1) -
\Omega_i^m(t))).$$ The condition to be fulfilled is
$$\Omega_i^n(t+1) - \Omega_i^m(t)\geq 0$$ that corresponds to the
imposed constraint.

\vspace{1cm}
It is important to note that in all the  cases, the
monocultural state corresponds  to the case when the Lyapunov
function takes an absolute minimum value, $\pounds_i^M$. We can
use this value for normalization, such that $L_i=\pounds_i /
\pounds_i^M$.

\begin{figure}[hbt]
\centering \resizebox{\columnwidth}{!} {\rotatebox[origin=c]{0}{
\includegraphics{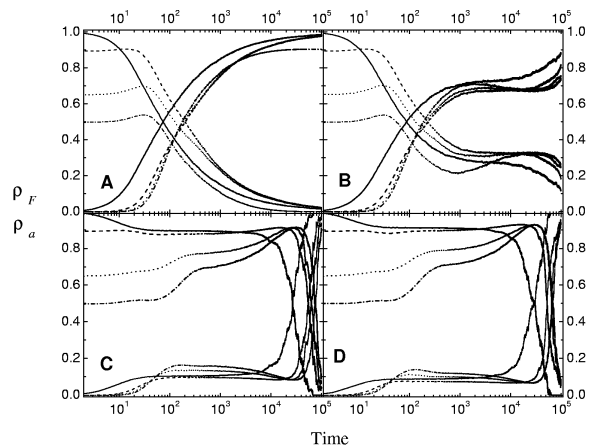}}}
\caption{Proportion of active links $\rho_a$ and of complete
overlap links $\rho_F$ vs. time, with $q=2$ (full), $q=5$
(dashed), $q=10$ (dotted), $q=15$ (dotted-dashed). Each plot
correspond to a different value of $p$: A)$p=0$, B)$p=0.01$,
C)$p=0.5$, D)$p=0.9$. Axelrod' s Case.} \label{vec00}
\end{figure}

\section{Numerical results}

In what follows we will include results corresponding to the cases
1 and 2 as well as those corresponding to the Axelrod' s original
model, for which we have not defined a Lyapunov function.

We have performed extensive numerical simulations of the described
model, considering different dynamics. The networks have $N=10^4$
vertices and connectivity $K=2$. A typical realization starts with
the generation of the random network and the initialization of the
state of the elements. After a transient period, the duration of
which depends on the parameters of the particular simulation, a
macroscopic stationary state is achieved. The computations are
then repeated for several thousand time steps to perform
statistical averages. We consider that the system has achieved a
macroscopic stationary state when the corresponding Lyapunov
function of the systems reaches a  stationary value. We will see
that this does not imply that the system is steady in a particular
microscopical state. Indeed, the configuration of the system
fluctuates among states with equal value of the Lyapunov function.
There are several aspects characterizing the evolution of the
system to a stationary value of the Lyapunov function. We also
analyze the behavior of the system subject to the original
dynamics, when the steady state is not characterized by a Lyapunov
function but by a microscopically frozen state.

In the calculations whose results are described in what follows,
we took $F=10$ and several values of $q$, ranging from 2 to 15. At
each time step only one change was proposed so the system was
updated asynchronically. We considered that one unity of time
corresponded to $N$ time steps.

For each of the dynamics described above, we have analyzed several
aspects of the evolution of the system. First we have calculated
the proportion of overlaps $\rho$ between individuals
corresponding to three cases, a) $\rho_0$ when $\omega_{(i,j)}=0$,
b)$\rho_F$ when $\omega_{(i,j)}=F$ and c)$\rho_a$ when
$0<\omega_{(i,j)}<F$. The cases a) and b) correspond to the
situation when no change in the system is possible because the
interaction of two individuals: in the case a) because no
interaction will occur when the cultural overlap is equal to zero,
in the case b) because individuals are already culturally
identical. The only active links are those corresponding to he
case c). Then we have calculated the corresponding Lyapunov
function (when defined) to show how its value evolves
monotonically to a steady one. Though this does not provide any
information about the inner structure of the system, or about the
existence of clusters, it helps us to have an idea of the amount
of cultural differentiation that is present. To show that though
the Lyapunov function reaches a steady value, but the system is
not in a steady state, we calculated the amount of changes that
occur in each time step. This was also useful to show that in the
Axelrod case, the system attained a frozen state, with no changes.

\subsection{Axelrod's Case}
This case corresponds to the original model
\cite{axel,palm2,palm1} where individuals interact with only one
of their neighbors at each time step. The interaction is mediated
by the cultural affinity, defined through the cultural overlap
$\omega_{(i,j)}$. The stronger the affinity is, the greater the
possibility of interaction between two subjects. In the present
work, the individuals are located on networks with different
degrees of disorder. The ordered case $p=0$, corresponds to a
one-dimensional lattice with interactions between the first and
second neighbors.
\begin{figure}[hbt]
\centering \resizebox{\columnwidth}{!} {\rotatebox[origin=c]{0}{
\includegraphics{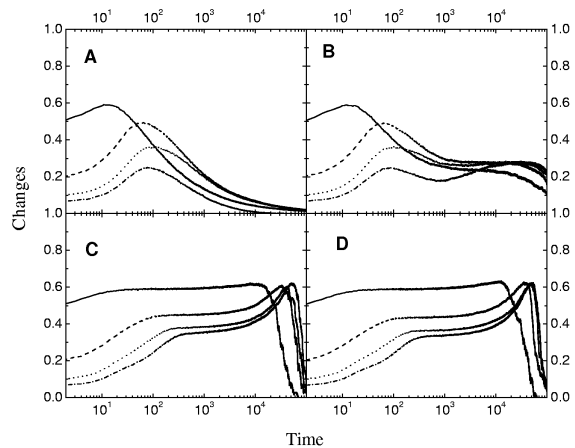}}}
\caption{Proportion of changes vs. time, for different values of
$q$ and $p$ as in Figure 1. Axelrod's Case.} \label{c00}
\end{figure}
As stated before, we did not find a Lyapunov function for this
case, and we restrict the displayed results to the time dependence
of the proportion of overlaps $\rho_F$ and $\rho_a$, and of the
proportion of  changes in the individuals' cultural profiles. We
recall that $\rho_0=1-\rho_F+\rho_a$. Figure \ref{vec00} displays
the time evolution, averaged over 1000 realizations, of the values
$\rho_F$ and $\rho_a$, corresponding to the amount of overlaps
$\omega_{(i,j)}=F$ and $0<\omega_{(i,j)}<F$, normalized to the
total number of links, $KN$. We evaluate these quantities on
networks with different degree of disorder, namely, $p=0$, 0.01,
0.5,0.9.
\begin{figure}[hbt] \centering \resizebox{\columnwidth}{!}
{\rotatebox[origin=c]{0}{
\includegraphics{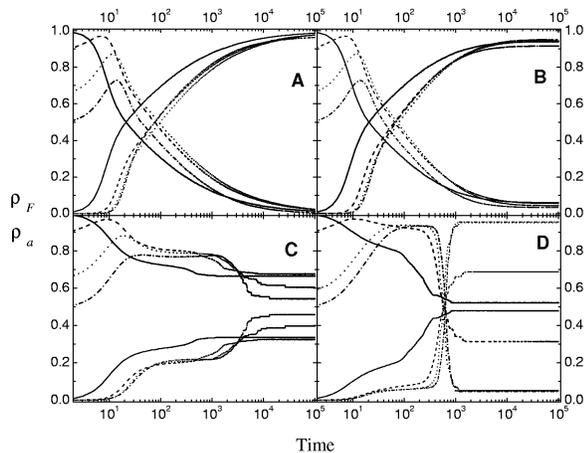}}}
\caption{Proportion of active links $\rho_a$ and of complete
overlap links $\rho_F$ vs. time, for different values of $q$ and
$p$ as in Figure 1. Case 1.} \label{vec10}
\end{figure}
A deeper insight into what is happening is obtained by analyzing
the data contained in Figure \ref{c00}. There we show the
proportion of changes in the cultural vectors of the individuals.
Each change corresponds to a component of any cultural vector that
changed its value. We show the amount of changes in a unit of time
normalized to the maximum value allowed $N$, the number of
proposed changes.

Figures 1.A and 2.A correspond to an ordered underlying network.
The system goes to a state where only non active links survive,
that is, $\rho_a\rightarrow0$. At the same time, while the system
reaches a steady state, associated with the number of changes
approaching 0, the system achieves  a monocultural state when
$q\leq F$ but goes to a multicultural state for higher values of
$q$. These facts have been already observed in \cite{axel,palm2}.

When some disorder is introduced into the network, the behavior of
the system is more complex. By looking at Fig.\ref{c00} we can see
that there are two different behaviors for ordered and very
disordered networks while the intermediate case, $p=0.01$ shows a
mixture of both. The system, in ordered networks evolves rather
fast to a state of low multiculturality or monoculturality. When
the disorder is increased, the initial disorder survives for
longer times. At the end, the system ends in a monocultural state
with except when $p=0$ and $q>F$. Figure \ref{vec00}, showing the
number of changes in time confirms what was mentioned before.
\begin{figure}[hbt]
\centering \resizebox{\columnwidth}{!} {\rotatebox[origin=c]{0}{
\includegraphics{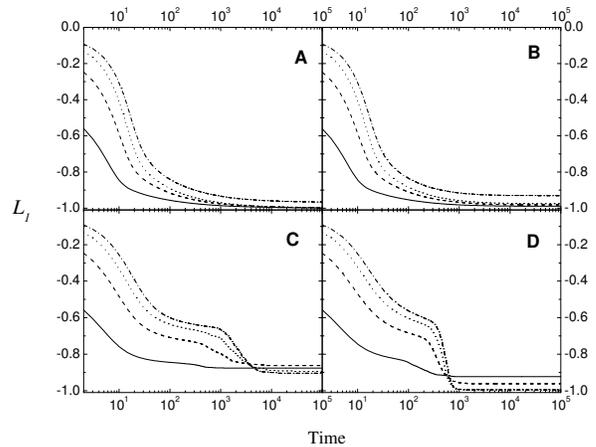}}}
\caption{Normalized Lyapunov function $L_1$ vs. time, for
different values of $q$ and $p$ as in Figure 1.} \label{l10}
\end{figure}
We have not observed sharp transitions while varying the $p$
value, the behavior of the system indeed, undergoes a smooth
change as the spatial disorder is increased.

\subsection{Case 1: Restricted cultural affinity}
\begin{figure}[hbt]
\centering \resizebox{\columnwidth}{!} {\rotatebox[origin=c]{0}{
\includegraphics{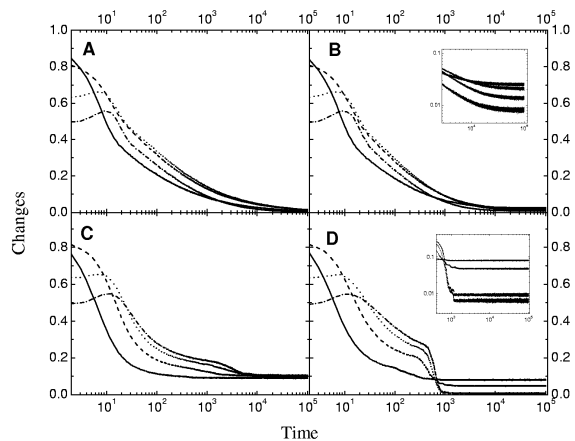}}}
\caption{Proportion of changes vs. time, for different values of
$q$ and $p$ as in Figure 1. Case 1.} \label{c10}
\end{figure}

In the following cases the calculation of the Lyapunov function
will provide us additional information about the system behavior.
As in the previous case, Figure \ref{vec10} displays the time
evolution, averaged over 1000 realizations, of the values $\rho_F$
and $\rho_a$, evaluating these quantities on networks of varying
disorder. Figure \ref{c10} shows the evolution in time of the
proportion of changes. The evolution of the normalized Lyapunov
function corresponding to this case, $L_1$ is plotted in Figure
\ref{l10}. Starting from the ordered case, Figure 3.A, we observe
that results do not differ so much from what we have previously
seen. Again, the system goes to a state where only non active
links survive, reaching a steady state,and achieving
monoculturality when $q\leq F$ and a certain degree of
multiculturality  for higher values of $q$.
This time we can recur to the Lyapunov function to see that the
absolute minimum is reached when $q\leq F$, but the system remains
in a frozen state of multiculturality when $q > F$. It is
interesting to observe that the number of changes, Figure
\ref{c10}.A,  goes to zero.
\begin{figure}[hbt]
\centering \resizebox{\columnwidth}{!} {\rotatebox[origin=c]{0}{
\includegraphics{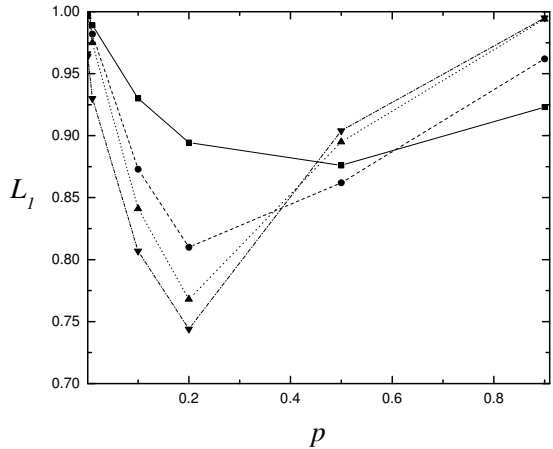}}}
\caption{Steady value of $L_1$ vs. $p$, with $q=2$ (full), $q=5$
(dashed), $q=10$ (dotted), $q=15$ (dotted-dashed).} \label{l1vsp}
\end{figure}
When disorder is included on the network, the behavior of the
system displays non trivial effects as can be can observed in
figures 3.B, 3.C and 3.D. The number of active links is different
from zero, even when a a steady value for the Lyapunov function is
reached. Though the monocultural state is the absolute minimum, it
is note attained by the system, who finishes trapped in local
minimum.  In Figures \ref{l10}.B, .C and .D we observe that the
Lyapunov function decreases monotonically to attain a steady state
but not to the absolute minimum. On the other hand, the steady
values depend non monotonically on the disorder of the network.
Despite the fact that $\pounds_1$ remains steady, the state of the
system is not frozen. This affirmation comes from the observation
of Figure \ref{c10}, where we find that the number of changes
remains above zero in all cases. Again, the mean value of changes
behaves in a non trivial way when $q$ or $p$ change.

Perhaps the most interesting feature is the interplay between the
effect of the spatial disorder and the values of $q$. This can be
better observed by analyzing the behavior of the Lyapunov
function. In some cases the disorder introduced by the network
prevents the system from achieving the previously reached
monocultural state, but on the other hand, the final degree of
multiculturality depends in a very interesting way from both
parameters. A new concept should be used now, the concept of
plasticity of the system. We can link this concept to the amount
of changes that occur. The plasticity grows with $p$ but then
decays again.  A similar non monotonic behavior in terms of $p$
can be observed for $L_1$, plotted in Figure \ref{l1vsp}.
\begin{figure}[hbt]
\centering \resizebox{\columnwidth}{!} {\rotatebox[origin=c]{0}{
\includegraphics{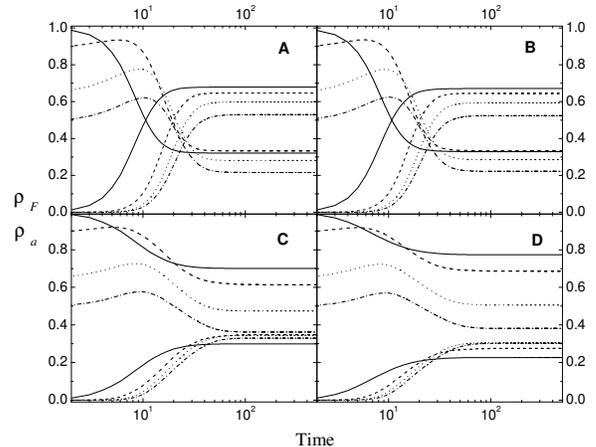}}}
\caption{Proportion of active links $\rho_a$ and of complete
overlap links $\rho_F$ vs. time, for different values of $q$ and
$p$ as in Figure 1. Case 3.} \label{vec30}
\end{figure}

\subsection{Case 2: Complete Cultural affinity }
The first aspect that we can observe for this case is that
independently of the degree of disorder of the network, the state
of monoculturality is never achieved, as shown in  Figure
\ref{vec30}. We can again verify the interplay between the
parameters $q$ and $p$ and their effect on the behavior of the
system. Another issue to be observed is the time scale. The
evolution towards a steady value of the Lyapunov function is much
faster than before. At the same time we observe that by increasing
the disorder the amount of active links grows. \begin{figure}[hbt]
\centering \resizebox{\columnwidth}{!} {\rotatebox[origin=c]{0}{
\includegraphics{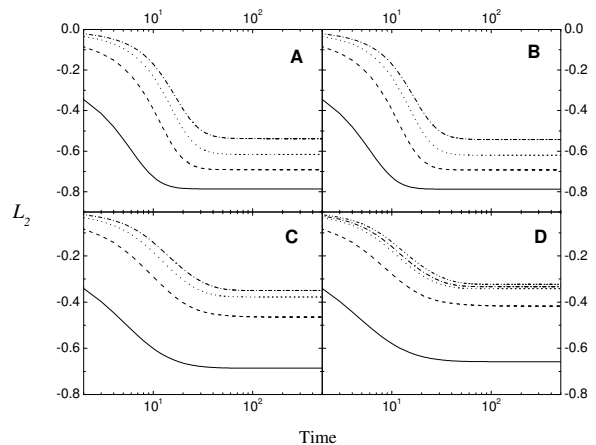}}}
\caption{Normalized Lyapunov function $L_2$ vs. time, for
different values of $q$ and $p$ as in Figure 1.} \label{l30}
\end{figure}
This is associated to the fact observed in Figure \ref{c30}, where
we can see how the amount of changes in the final state also
increases with $p$. The behavior of the Lyapunov function simply
verifies that the system reaches a steady value and that this
value is far from being the absolute minimum. In all the cases,
the steady value decreases with $q$.

\begin{figure}[hbt]
\centering \resizebox{\columnwidth}{!} {\rotatebox[origin=c]{0}{
\includegraphics{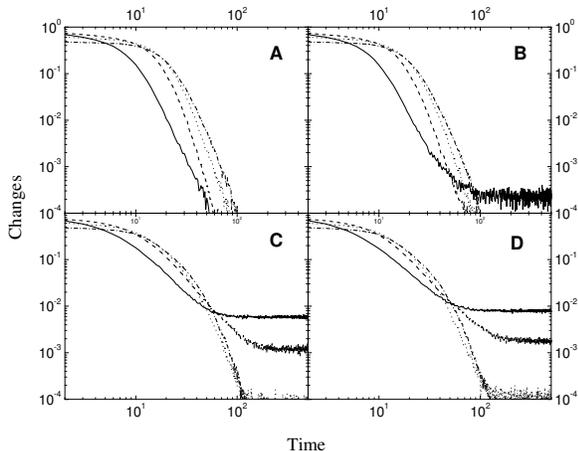}}}
\caption{Proportion of changes vs. time, for different values of
$q$ and $p$ as in Figure 1. Case 2.} \label{c30}
\end{figure}

\section{Conclusions}

Axelrod's model shows how a microscopical local process of
interaction, leading to convergence provokes the emergence of
global polarization. In previous works, the model was used to
analyze the effect of the number of cultural aspects and traits on
the steady configuration of the system. Further analysis
\cite{cast} of the relative size of the largest cultural domain
revealed an order disorder transition with $q$, the number of
different traits, playing the role of the control parameter. Under
a threshold value $q_c(F)$,the system converges to a monocultural
uniform state. Above $q_c(F)$ the system freezes in a
multicultural state, that can be associated to polarization. The
stability of the multicultural states was analyzed in  \cite
{palm2} by perturbing the system when frozen in a multicultural
state and showing the further convergence to the monocultural
sate. Perturbations were associated to cultural drift.

In this work we proposed a different sort of generalization of
Axelrod' s model. We modified the model to include interactions
among several individuals within a neighborhood or to let each
individual evaluate the changes in its cultural preferences by
analyzing those of its neighbors. Different ways of considering
this extended interaction were shown. For each, an associated
Lyapunov function was found, letting us analyze the convergence of
the system towards an absolute or local minima. The disorder of
the system was not reduced to that introduced by the initial
condition by increasing the value of $q$, but also included in the
spatial distribution of the agents. For this purpose we analyzed
the effect of the disorder of the underlying network considering
small world networks of varying disorder.

The results should be analyzed from several different points of
view. As expressed in the introduction, we not only include
initial cultural disorder through higher values of $q$, but also
spatial disorder within the underlying network. The results linked
to this aspects can be compared with previous results and thus
unveil the effect of the newly defined interaction of each
individual with the whole neighborhood. As already known,
increasing the value of $q$ leads the system to undergo a
transition from monoculturality to multiculturality. However, the
effect of spatial disorder attempts against this effect. This can
be explain by recalling that in a disordered network the
clusterization of the system is lower and thus, the existence of
clusters of culture reflected in a polarized situation is no
longer achieved. When the slightest disorder is added to the
network, the number of links with overlap equal to zero goes to
zero.

The change of the rules of interaction introduces a new
interesting behavior. Not only does the amount of active links not
go to zero, with the exception when the underlying lattice is
ordered and $q=2$, but also the system reaches a situation when
the Lyapunov function adopts a steady value but the system is not
frozen. The configuration of the system changes in time, as can be
observed from the figures displaying the number of changes in
time.

The results presented here complement what was already found in
the analysis of the model first proposed by Axelrod. The
interesting feature is that the system, despite reaching a steady
situation, does not remains static. Some aspects still deserve
further analysis. Among them we will consider in a future work the
inclusion of noise.

\end{document}